\documentclass[aps,prd,reprint,twocolumn,preprintnumbers,superscriptaddress,nofootinbib]{revtex4-2}

\usepackage{amsmath,amssymb,bm,bbm,amsfonts}
\usepackage{physics}
\usepackage{graphicx,graphics}
\usepackage[dvipsnames]{xcolor}
\usepackage[colorlinks=true,linkcolor=blue,citecolor=blue,urlcolor=blue]{hyperref}
\usepackage{dsfont}
\usepackage{comment}
\usepackage{hhline,multirow}
\usepackage{dcolumn}
\usepackage{url}
\usepackage[normalem]{ulem}
\usepackage{mathrsfs}
\usepackage{latexsym}
\usepackage{amscd}
\usepackage{slashed}
\usepackage{mathtools}
\usepackage{array}
\usepackage{orcidlink}

\newcommand{\jlab}{Thomas Jefferson National Accelerator Facility, Newport  News, Virginia 23606, USA}
\newcommand{\indiana}{Department of Physics, Indiana  University, Bloomington, Indiana 47405, USA}

\newcommand{\wm}{Department of Physics, William \& Mary, Williamsburg, Virginia 23185, USA}

\begin{document}

\title{Variational Neural Network Approach to QFT in the Field Basis}

\author{K.~\surname{Braga}\orcidlink{0009-0008-8752-8292}}
\email{kmbraga@wm.edu}
\affiliation{\wm}

\author{N.~\surname{Sato}\orcidlink{0000-0002-1535-6208}}
\email{nsato@jlab.org}
\affiliation{\jlab}

\author{A.~P.~\surname{Szczepaniak}\orcidlink{0000-0002-4156-5492}}
\email{aszczepa@indiana.edu}
\affiliation{\jlab}
\affiliation{\indiana}

\begin{abstract}
We present a variational neural network approach for solving quantum field theories in the field basis, focusing on the free Klein–Gordon model formulated in momentum space. While recent studies have explored neural-network-based variational methods for scalar field theory in position space, a systematic benchmark of the analytically solvable Klein–Gordon ground state—particularly in the momentum-space field basis—has been lacking. In this work, we represent the ground-state wavefunctional as a neural network defined on a discretized set of field configurations and train it by minimizing the Hamiltonian expectation value. This framework enables direct comparison to exact analytic results for a range of key observables, including the ground-state energy, two-point correlators, expectation value of the field, and the structure of the learned wavefunctional itself. Our results provide quantitative diagnostics of accuracy and demonstrate the suitability of momentum space for benchmarking neural network approaches, while establishing a foundation for future extensions to interacting models and position-space formulations.
\end{abstract}

\preprint{JLAB-THY-25-4428}
\date{\today}
\maketitle

\tableofcontents

\section{Introduction}

Variational methods offer a nonperturbative route to quantum field theory (QFT) by approximating the ground state through an optimized trial wavefunctional.
 Applying this principle to QFT, however, has long been known to be exceedingly difficult—as Feynman noted, the infinite degrees of freedom,  ultraviolet divergences and the non-trivial nature of the vacuum pose serious challenges to constructing useful trial states~\cite{Feynman1988}.
 Early attempts like the Gaussian effective action and time-dependent Hartree-Fock exemplified the variational approach in field theory~\cite{Jackiw1979}. The application of variational many-body methods to QCD, specifically to address dynamical chiral symmetry breaking, the phenomenology of confinement, and the role of glue in generating the hadron spectrum, has been pursued in recent years~\cite{Greensite:2014bua,Greensite:2011pj,Kogan:1994wf,Szczepaniak:2003ve,Feuchter:2004mk,Reinhardt:2004mm}. Further progress will depend on using more flexible wave functions that deep neural networks can potentially provide. 

Recent advances in machine learning have revitalized this outlook. Neural-network quantum states have emerged as powerful variational \textit{ans\"atze} for many-body systems~\cite{Carleo2017}. Carleo and Troyer’s landmark work demonstrated that a neural network can efficiently encode the ground state of an interacting spin system, capturing entanglement and long-range correlations beyond traditional trial wavefunctions. Since then, neural network \textit{ans\"atze} have achieved compelling successes across quantum physics. Notably, deep networks have been used to obtain near-exact solutions of the electronic Schr\"odinger equation in atoms and molecules~\cite{Pfau2020,Hermann2020}, outperforming even “gold standard” methods by compactly representing high-dimensional continuous wavefunctions. Neural quantum states have also been applied to continuum many-body systems—for example, representing the superfluid wavefunction of an ultracold Fermi gas~\cite{kim2024neural} with higher accuracy than diffusion Monte Carlo.

Encouraged by these achievements, several groups have begun to tackle quantum field theories using neural networks. On the lattice side, Chen \textit{et al.}~\cite{Chen:2022asj} introduced flow-based neural wavefunctions to simulate $2+1$ dimensional lattice gauge theory, incorporating exact gauge symmetries and addressing finite-density sign problems. On the continuum side, Martyn \textit{et al.}~\cite{Martyn2022} introduced a Deep Sets neural network architecture to variationally approximate ground states of a nonrelativistic bosonic field theory in the Fock basis, successfully capturing a broad sector of the continuum many-body Hilbert space. Rovira \textit{et al.}~\cite{Rovira} investigated a feed-forward neural network for scalar $\phi^4$ theory on a discretized spatial lattice. These studies highlight the promise of neural networks in field theory, while underscoring key trade-offs: one must typically either truncate Fock space or discretize real space, each of which may obscure continuum physics.

Parallel efforts with continuous tensor networks, such as cMPS~\cite{Verstraete2010,Haegeman2013} and cPEPS~\cite{Shachar2022}, offer elegant continuum variational frameworks, but face challenges when scaling beyond 1D or incorporating general interactions. 
See Martyn \textit{et al.}~\cite{Martyn2022} for a more detailed discussion regarding further difficulties of incorporating matrix product states in the context of our work. This motivates exploring more flexible and numerically driven representations, like neural networks, that can be trained directly using variational principle.

A central challenge in formulating quantum state representations for field theories using neural networks lies in mapping a field configuration, defined over a continuous domain (e.g., position or momentum space), to a single numerical value (c-number). Because such configurations span an infinite-dimensional space, exact mappings are computationally infeasible. In practice, the problem becomes tractable by evaluating field configurations at discrete points, analogous to classical lattice methods. This discretization effectively reduces the continuum field theory to a high-dimensional quantum mechanical problem.

An equally important challenge is the representation and numerical implementation of field operators that are fundamentally defined in the continuum. In the field basis, the canonical momentum operator is expressed as a variational (functional) derivative with respect to the field configuration. Upon discretization, this derivative must be defined with explicit attention to grid spacing, in order to maintain correspondence with the continuum theory, a convention followed in all field-basis approaches, including recent studies such as Rovira \textit{et al.}~\cite{Rovira}, who set the grid spacing to unity for simplicity. In contrast, Fock basis representations, which work in terms of occupation numbers, rapidly become intractable or conceptually inadequate for strongly interacting or nonperturbative theories. The field basis, on the other hand, naturally accommodates arbitrary field configurations and is uniquely suited for describing nonperturbative quantum states where particle number is not a useful concept.

While our approach does not solve the field theory in the continuum, it maintains conceptual and operational alignment with continuum operator definitions at the discretized level. By employing a field-basis discretization consistent with the underlying theory, we benchmark our method against analytic solutions of the discretized Klein–Gordon model, computed with the same grid spacing and ultraviolet cutoff as used in our simulations. This ensures all comparisons are performed in a controlled, finite setting where exact results are available.

In this work, we present a neural-network-based variational method formulated in the Schr\"odinger
 picture using discretized field configurations. Each configuration is represented by constant field values within discrete Riemannian intervals, serving as variational input. Canonical operators like $\pi(x) = -i,\delta/\delta \phi(x)$ are retained in their continuum-inspired form and evaluated numerically via finite-difference approximations. The wavefunctional $\Psi[\phi(x)]$ is parameterized by a neural network trained to minimize the expectation value of the Hamiltonian.

Our numerical implementation discretizes momentum space to evaluate Riemann sums, introducing a finite momentum cutoff as the primary regulator, a standard practice for ultraviolet safe continuum QFT. Although we do not explicitly perform a continuum limit here, our method maintains the interpretability and operator definitions associated with canonical continuum field theory. We benchmark this approach on the free Klein–Gordon model in 1D, a theory with an analytically known ground-state wavefunctional. Working in the Schr\"odinger picture and focusing on stationary states, our formulation is inherently time-independent. The variationally learned wavefunctional accurately reproduces analytic ground-state observables, including the vacuum energy, mode-by-mode correlators, expectation values of the field, and the structure of the wavefunctional itself, enabling direct visualization of the vacuum structure.

By working in the field basis, our framework provides a practical and physically motivated route to applying machine learning to quantum field theory without requiring Fock-space truncations or path integrals. This approach is particularly well-suited for exploring nonperturbative and strongly coupled dynamics, and generalizes naturally to interacting scalar theories and, with further development, to gauge theories. Leveraging the flexibility of neural networks while retaining fidelity to the structure of canonical continuum theory, this work establishes a controlled foundation for variational studies of quantum field theory.

The remainder of this document is structured as follows. In Section II, we review the Schr\"odinger picture formulation of scalar QFT and outline the analytic structure of the Klein–Gordon ground-state wavefunctional. Section III describes our variational framework, including the neural network architecture, operator definitions, and discretization strategy. In Section IV, we benchmark our method against analytical solutions. Section V discusses the advantages, limitations, and future directions of our approach, placing it in context with other variational methods. Finally, Section VI provides a summary and outlook for future research.

\section{Theoretical Framework}
\label{sec:background}

As a proof of concept for solving quantum field theories in the field basis, we consider the free Klein-Gordon model. This case provides an analytically tractable benchmark and introduces many of the essential tools and operator structures relevant to more complex systems.
In the Schrödinger picture, the quantum states of the theory are represented as wavefunctionals $\Psi[\phi(x)]$ over field configurations $\phi(x)$. The Klein-Gordon Hamiltonian operator acting in position space  takes the form
\begin{equation}
H = \frac{1}{2} \int dx \left( -\frac{\delta^2}{\delta \phi(x)^2} + |\nabla \phi(x)|^2 + m^2 \phi(x)^2 \right),
\label{eq:position_hamiltonian}
\end{equation}
where the momentum operator is represented by the functional derivative $\pi(x) = -i\,\delta/\delta \phi(x)$. To facilitate numerical and analytical benchmarking, we work in momentum space where the free Hamiltonian diagonalizes and takes the form
\begin{equation}
H = \frac{1}{2} \int \frac{dk}{2\pi} \left( -\frac{\delta^2}{\delta \tilde{\phi}(k)^2} + (k^2 + m^2)\tilde{\phi}(k)^2 \right)\;.
\label{eq:momentum_hamiltonian}
\end{equation}
The momentum-space representation plays a central role in our numerical formulation. We simplify our model by enforcing $\tilde{\phi}(k) = \tilde{\phi}(-k)$. This allows us to restrict the domain of computation for computational simplicity.

To numerically solve the time-independent Schrödinger equation with the Hamiltonian in Eq.~(\eqref{eq:momentum_hamiltonian}), we discretize the momentum space over a finite interval $[0, k_{\text{max}}]$ into $N_k$ uniformly spaced grid points with spacing $\Delta k$. The field $\tilde{\phi}(k)$ is evaluated at these discrete momenta, resulting in a set of real-valued degrees of freedom $\tilde{\phi}_k$.
Replacing the momentum integral with a Riemann sum, the discretized Hamiltonian becomes
\begin{equation}
H = \frac{1}{2} \sum_{k=1}^{N_k} \frac{\Delta k}{2\pi} \left( -\frac{\delta^2}{\delta \tilde{\phi}_k^2} + (k^2 + m^2)\tilde{\phi}_k^2 \right)\;,
\label{eq:discrete_hamiltonian}
\end{equation}
where $k$ indexes the discrete momentum modes. The functional derivatives acting on the wavefunctional can be estimated using a standard finite difference approach \cite{}. Specifically, the first functional is computed via
\begin{equation}
\frac{\delta \Psi[\tilde{\phi}]}{\delta \tilde{\phi}(k)} = \lim_{\epsilon \to 0} \frac{\Psi[\tilde{\phi} + \epsilon\,\delta_k] - \Psi[\tilde{\phi}]}{\epsilon},
\end{equation}
where $\delta_k$ denotes a localized variation in the $k$-th momentum mode. In the continuous limit, it becomes the Dirac Delta function. Similarly, the second variational derivative follows from applying this definition twice.

The ground-state wavefunctional, is known analytically for the free theory~\cite{Hatfield}, and takes a factorized from 
\begin{equation}
\Psi_0[\tilde{\phi}] \propto \exp\left( -\frac{1}{2} \sum_{k=1}^{N_k} \frac{\Delta k}{2\pi} \, \omega_k \, \tilde{\phi}_k^2 \right)\;.
\label{eq:discrete_groundstate}
\end{equation}
Here, $\omega_k = \sqrt{k^2 + m^2}$ denotes the energy of the momentum mode $k$. The ground state wavefunctional in the discretized theory corresponds to a system of independent, decoupled harmonic oscillators in momentum space. The corresponding ground-state energy is obtained by summing up the zero-point energies across all modes, 
\begin{equation}
E_0 = \frac{1}{2} \sum_{k=1}^{N_k} \omega_k\;.
\label{eq:ground_energy}
\end{equation}
While this quantity is divergent in the continuum, it is well-defined for a theory with a  finite cutoff $k_{\text{max}}$, and forms a useful variational benchmark for evaluating the accuracy of learned wavefunctionals. 

\section{Variational Method with Neural Wavefunctionals}

\begin{figure}[t]
\centering
\includegraphics[width=0.45\textwidth]{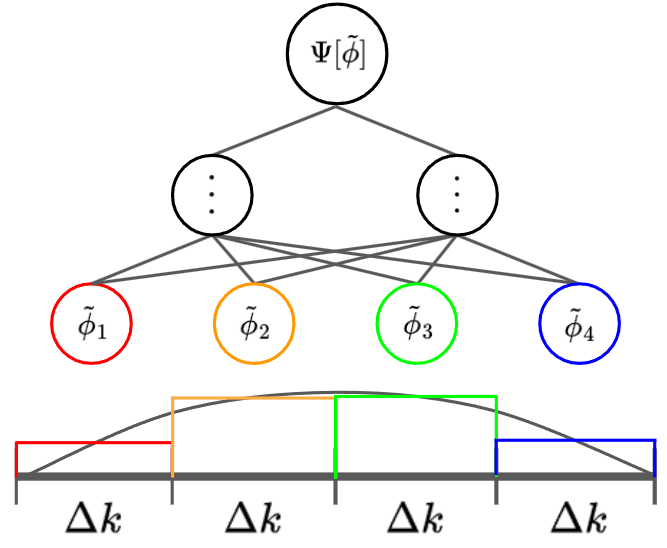}
\caption{\textbf{Top}: schematic architecture of the neural network representing the wavefunctional, which maps field configurations to amplitudes. \textbf{Bottom}: discretization of a continuous momentum-space field configuration, which serves as input to the network.}
\label{fig:architecture}
\end{figure}

In this section, we present a machine learning framework  to approximate the ground state wavefunctional of a free scalar field theory. We use  a neural network  to parameterize the wavefunctional in momentum space, and it is trained to minimize the expectation value of the Hamiltonian in Eq.~(\ref{eq:discrete_hamiltonian}). 
As mentioned, we work in momentum space and discretize the domain $k \in [0,k_{\max}]$ uniformly with spacing $\Delta k$, yielding $N_k$ total grid points. 
A single field configuration is represented as
$$
\tilde{\phi} \equiv (\tilde{\phi}_1, \tilde{\phi}_2, \dots, \tilde{\phi}_{N_k}),
$$
with components corresponding to momentum modes of the grid. A schematic representation of the wavefunctional is shown in Fig.~\ref{fig:architecture}

To evaluate the Hamiltonian on a field configuration $\tilde{\phi}$, we approximate the second variational derivative in Eq.~(\ref{eq:discrete_hamiltonian}) using a five-point finite difference stencil, 
\begin{align}
\frac{\delta^2 \Psi[\tilde{\phi}]}{\delta \tilde{\phi}_k^2} 
&\approx \frac{1}{12\epsilon^2} \Big( -\Psi[\tilde{\phi}+2\epsilon \delta_k] + 16\Psi[\tilde{\phi}+\epsilon \delta_k] \nonumber \\
&\quad - 30\Psi[\tilde{\phi}] + 16\Psi[\tilde{\phi}-\epsilon \delta_k] - \Psi[\tilde{\phi}-2\epsilon \delta_k] \Big)\;.
\end{align}
Here, $\delta_k = 2\pi/\Delta k$ denotes a perturbation at momentum index $k$, and $\epsilon$ is a hypermeter for the optimization problem. 

The expectation value of the Hamiltonian in Eq.~(\ref{eq:discrete_hamiltonian})  can be estimated using Monte Carlo (MC) methods. Specifically, we generate field configurations according to a probability density $p(\tilde{\phi})$ given in terms of the wavefunctional, 
\begin{align}
\tilde{\phi} \sim p(\tilde{\phi})\propto|\Psi(\tilde{\phi})|^2\;.
\label{eq:phi-density}
\end{align}
To sample the density, we use the \texttt{vegas} algorithm, which implements importance sampling. Field configuration samples are then generated after performing adaptive integration of the density using \texttt{vegas} code. With a giving a set of $N_f$ generated samples, we estimate the expectation value of the Hamiltonian as
\begin{equation}
\langle \hat{H} \rangle \approx \frac{\sum_{i=1}^{N_f} \Psi[\tilde{\phi}^{(i)}] \, \hat{H} \Psi[\tilde{\phi}^{(i)}]}{\sum_{i=1}^{N_f} \Psi^2[\tilde{\phi}^{(i)}]}\;.
\label{eq:discexpvalHam}
\end{equation}
As it is standard in variational approach, the neural netework is optimized to minimize Eq.~(\ref{eq:discexpvalHam}). For each epoch of the training, we generate new field configurations with the updated wavefunctional  We avoid the use of mini-batching as we found that small subsets of field configurations provide insufficient coverage of the field configuration space and degrade convergence. Instead, we use the full set of generated field configurations to updated the parameters of the neural network.

Once the training of the neural network wavefunctional is completed, additional observables can be estimated via MC methods in a manner similar to Eq.~(\ref{eq:discexpvalHam}). In particular, we estimate the expectation value of the fields and the two-point correlation functions, given respectively by
\begin{equation}
\left\langle \tilde{\phi}_k \right\rangle
\approx 
\frac{\sum_{i=1}^{N_f} \tilde{\phi}_k^{(i)} 
\Psi^2[\tilde{\phi}^{(i)}]}{\sum_{i=1}^{N_f} \Psi^2[\tilde{\phi}^{(i)}]}.
\label{eq:meanfield}
\end{equation}
\begin{equation}
\left\langle \tilde{\phi}_j \tilde{\phi}_{k} \right\rangle
\approx 
\frac{\sum_{i=1}^{N_f} \tilde{\phi}_j^{(i)} \tilde{\phi}_{k}^{(i)} \Psi^2[\tilde{\phi}^{(i)}]}{\sum_{i=1}^{N_f} \Psi^2[\tilde{\phi}^{(i)}]}.
\label{eq:twopoint}
\end{equation}
In order to access the convergence of our observables, we report the variance across 100 independent MC estimate for each observable.

\section{Computational setup}
\label{sec:setup}

Our variational analysis is carried out using $N_k = 8$ discretized momentum-space points, uniformly spaced in the range $0<k<1$. The field configurations $\tilde{\phi}_k$ are constrained such that $|\tilde{\phi}_k| \leq A$, where $A$ is chosen large enough to encompass the physically relevant fluctuations. If $A$ is set too small, high-amplitude configurations are excluded from the training set, leading to a systematic underestimation of observables that are sensitive to the distribution tails. We find that $A \sim 20 $ provides sufficient coverage for our benchmark problem. Although larger values of $A$ expand the accessible phase space, they also increase both the dimensionality and steepness of the functional landscape, thereby increasing the complexity of the training process.

For each training epoch, we generate $N_f = 50{,}000$ independent field configurations to estimate the ground state energy in Eq.~(\ref{eq:discexpvalHam}). The neural network architecture consists of a simple feed-forward model with one hidden layer of 256 neurons—a ``$8 \times 256 \times 1$'' layout. This choice balances expressivity, computational cost, and training stability.

Finally, the stopping criteria for the model training is determined dynamically. After each epoch, we compute the relative standard deviation of $\langle \hat{H} \rangle$ over the last 50 recorded values. If this ratio falls below a threshold of 0.1\%, training is halted.

\section{Results}
\label{sec:results}

\begin{figure}[t]
\centering
\includegraphics[width=0.48\textwidth]{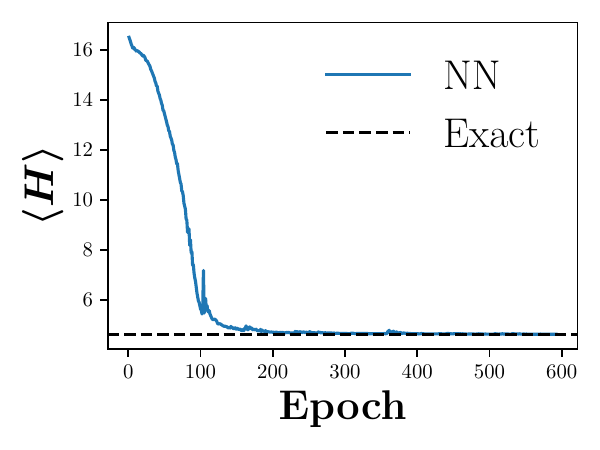}
\caption{Training history of the neural network energy estimate $\langle H \rangle$ as a function of training epoch. The solid line shows the network’s prediction; the dashed line indicates the exact ground-state energy. Rapid convergence is followed by minor fluctuations due to stochastic field sampling and training noise.}
\label{fig:training_curve}
\end{figure}

\begin{figure*}[t]
\centering
\includegraphics[width=\textwidth]{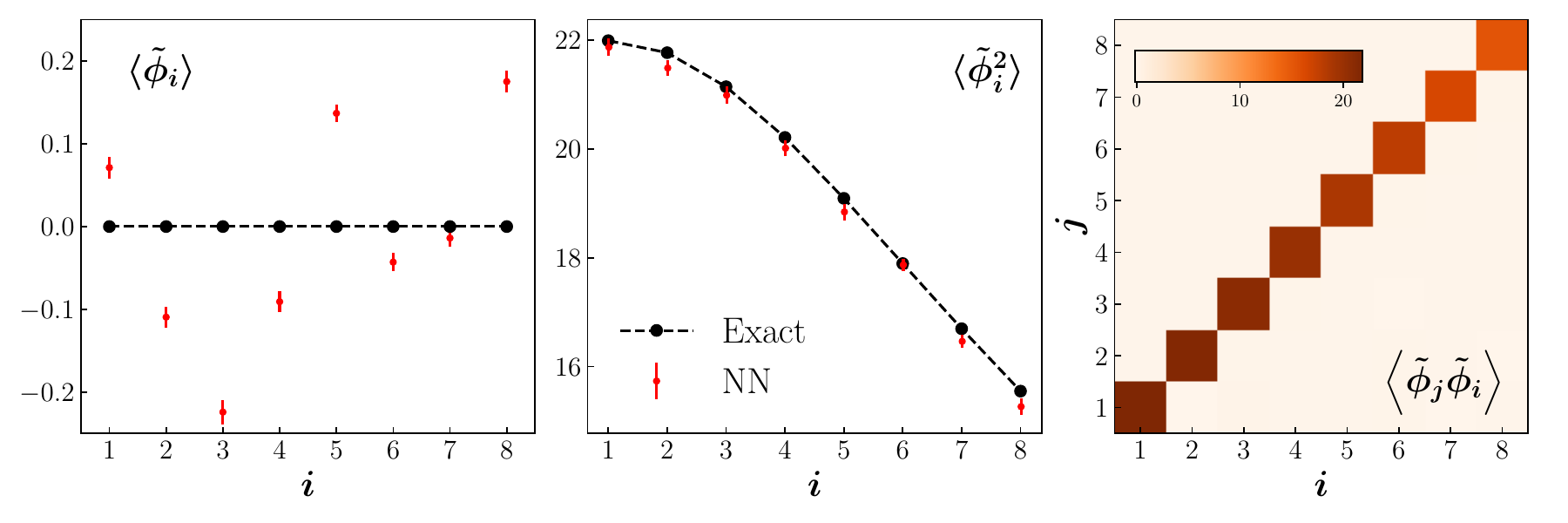}
\caption{Mean field configurations (left) and  the 
Two-point correlation matrix 
$ \langle \tilde{\phi}_i \tilde{\phi}_j \rangle$ (middle and right) of the Klein-Gordon field in momentum space.}
\label{fig:correlations}
\end{figure*}

We present in Fig.~\ref{fig:training_curve} the optimization of the neural network wavefunctional as a function of training epochs, illustrating how the ground state energy estimate, or cost function, converges toward the expected theoretical result for the Klein-Gordon Hamiltonian. After a rapid initial descent, the estimate plateaus and exhibits mild fluctuations, reflecting the noise introduced by stochastic field sampling. The reconstructed ground state energy from 100 independent MC evaluations is $4.6206 \pm 0.0060$ consistent within the uncertainties of the exact value of $4.6250$. This consistency  indicates that the learned wavefunctional not only converges during training, but also generalizes well to unseen sets of field configurations.

To assess whether the trained neural network captures the correct structure of the ground state wavefunctional beyond merely minimizing the energy, we present in Fig.\ref{fig:correlations} the MC estimates of the mean field configuration (left) and the two-point correlation functions (middle and right), computed using Eqs.(\ref{eq:meanfield},~\ref{eq:twopoint}). The uncertainties are estimated using as before 100 independent MC evaluations. As expected for a translationally invariant free theory, the mean field $\langle \tilde{\phi}(k) \rangle$ remains close to zero with fluctuations around the expected values. This observable is generally more challenging to reconstruct accurately, as it is sensitive to small asymmetries in the learned distribution and can be dominated by statistical noise due to its vanishing expectation value in the exact theory.

The diagonal terms of the two-point correlator are shown in the middle panel. While minor deviations from the exact results are present, the neural network wavefunctional captures the overall trend reasonably well. Similarly, the right panel shows the full correlator matrix, which, as expected for a free scalar field theory, displays a characteristic structure with prominent diagonal entries and negligible off-diagonal components. This qualitative structure is faithfully reproduced by the neural network.

The consistency of the vanishing mean field, the accurate reconstruction of the two-point correlators, and the convergence of the ground state energy together provide strong evidence that the trained neural network faithfully represents the true ground state wavefunctional of the theory. Moreover, it demonstrates the model's ability to generalize and predict physical observables not directly used during training.

\begin{figure*}[t]
\centering
\includegraphics[width=\textwidth]{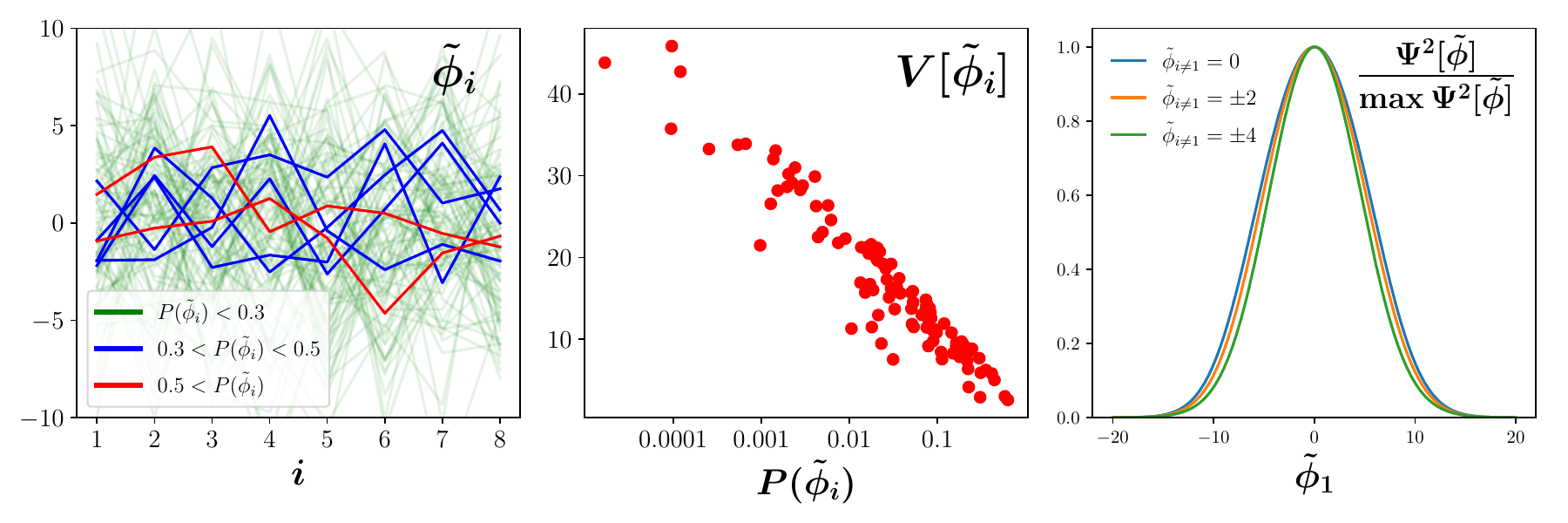}
\caption{MC field configurations sampled from the trained NN-based wavefunctional (left), variance of the field configurations across the discretized momentum space as a function of probability (middle), and squared wavefunctional as a function of $\tilde{\phi}_1$ under different fixed values of the remaining field components (right).}
\label{fig:fields}
\end{figure*}

An advantage  of our framework is the ability to directly visualize stationary states, such as the ground state wavefunctional. In many systems, the wavefunctional encodes essential information about how the field is distributed in its lowest-energy configuration. Visualizing this functional thus provides an insight into the structure and behavior of the vacuum. 

In Fig.~\ref{fig:fields}, we present such a visualization. In the left panel, we plot the MC field configurations across the momentum grid, grouping the samples into three ranges according to their associated probability values. Each vertex of the plotted curves corresponds to the center of the momentum interval. In the discretization of the quantum field theory, these intervals represent the finite-width subdivisions used to approximate integrals over continuous momenta. 

We observe that low-probability samples exhibit a significantly larger spread in the field configurations compared to those with higher probabilities. Specifically, samples with higher probabilities show reduced fluctuations and display smoother behavior, as is evident in the middle panel, where the variance of the field configurations across the 8 grid points is plotted as a function of probability. This trend is indeed expected because high-probability samples are more representative of the typical field configurations that reflect the dominant structure of the ground state. In contrast, low-probability samples arise from the tails of the distribution, where the neural network is less constrained by variational approach, and fluctuations are naturally amplified due to the decreased likelihood and increased statistical uncertainty.

Finally, in the right panel of Fig.~\ref{fig:fields}, we show the squared wavefunctional as a function of the field configuration at site $i=1$. The curves correspond to three different conditions, where the remaining fields $\tilde{\phi}_{i\neq1}$ are fixed to alternating values of $0$, $\pm 2$, and $\pm 4$ across the momentum grid. For instance, the configuration $\tilde{\phi}_{i\neq1} = \pm 2$ corresponds to $(\tilde{\phi}_1, 2, -2, 2, -2, 2, -2, 2)$. Each curve is normalized by its maximum value to highlight differences in shape.

The purpose of this plot is to demonstrate that, as expected theoretically, the neural network–based wavefunctional approximates the ground state of a system of decoupled harmonic oscillators. In such a system, the joint probability distribution factorizes across sites, so that aside from normalization, the shape of the squared wavefunctional is independent of the values of the other field components. That is, the probability distribution satisfies
\begin{equation}
P(\tilde{\phi}_{1:8}) = \prod_i P(\tilde{\phi}_i); .
\end{equation}
The consistency of the wavefunctional’s shape across different filed configurations provides strong evidence that the learned distribution correctly reflects the factorized structure of the true ground state.

To summarize, our visualizations reveal key features of the learned wavefunctional, such as smoothness, symmetry, and factorization, that are characteristic of the true ground state of a free theory. The wavefunctional visualization serves as an intuitive and complementary tool to traditional numerical diagnostics, offering insight into the structure of quantum states that might otherwise remain opaque. In particular, our results show that the neural network–based wavefunctional responds not merely to isolated mode displacements, but to the collective configuration of the entire field. These visualizations underscore the inherently global nature of the functional and shed light on the types of correlations and structures the network learns to encode.

\section{Discussion and Outlook}

In this work, we have introduced and benchmarked a variational method for solving the Klein-Gordon model in the field basis using neural networks. Our approach reconstructs the ground state wavefunctional and computes key observables including the energy, two-point correlator, and mean field.

These results establish a controlled and physically motivated framework for applying machine learning to quantum field theory. While demonstrated here for a free scalar field in one spatial dimension, the formalism extends naturally to interacting theories, multiple field species, and higher-dimensional systems. The ability to represent and visualize the full wavefunctional may be especially valuable in gaining intuition about field-theoretic ground states in more physically realistic models.

Several avenues for future development are evident. These include the treatment of complex-valued wavefunctionals, more efficient sampling strategies in high-dimensional field spaces, and architectural adaptations for incorporating symmetries or gauge constraints. Applications to benchmark models such as $\phi^4$ theory, the Schwinger model, or scalar QED could provide valuable insights and further test the scalability of the approach. Immediate next steps would involve solving the wave-functional in position space and  adding a $\phi^4$ interaction term, and extending to higher dimensions.

Another important extension involves the treatment of derivatives in both the field and wavefunctional spaces. While the current implementation operates in momentum space with explicitly discretized field configurations and finite-difference approximations, future formulations, especially in coordinate space, may benefit from treating these discrete values as samples from an underlying smooth field. This perspective enables the construction of interpolants whose derivatives can be used to evaluate both the field gradients appearing in the potential term and the second functional derivatives of the wavefunctional that define the kinetic term. In such a formulation, the entire Hamiltonian argument, rather than just isolated terms, would inherit a continuous structure, improving consistency between integration and differentiation and potentially enhancing numerical stability. This interpolation-based approach may serve as a continuum-consistent alternative to finite-difference stencils, particularly valuable in interacting theories or formulations involving nontrivial geometry.

Looking further ahead, this framework raises the intriguing possibility of directly visualizing nonperturbative phenomena in quantum field theory, including both topological solitons and bound states in QCD. As an aspirational example, one could imagine constructing the wavefunctional for the proton itself, not as a mere collection of Fock-space excitations, but as a stationary state of QCD. While realizing such a goal would require significant technical advancements, the conceptual groundwork laid here may contribute to that longer-term vision.

More broadly, field-space variational methods informed by neural representations may offer a complementary direction for nonperturbative studies in QFT. In particular, they provide a new means of accessing and interpreting the structure of stationary states, a perspective that could prove useful in understanding phenomena such as vacuum structure, confinement, and the emergence of bound states in gauge theories.

\begin{acknowledgments}
This work was supported by the DOE contract No.~DE-AC05-06OR23177, under which Jefferson Science Associates, LLC operates Jefferson Lab.  The work of KB and NS was supported by the DOE, Office of Science, Office of Nuclear Physics in the Early Career Program.
This work was supported by the DOE contract No.~DE-AC05-06OR23177, under which Jefferson Science Associates, LLC operates Jefferson Lab.  The work of KB and NS was supported by the DOE, Office of Science, Office of Nuclear Physics in the Early Career Program. APS acknowledges support from \mbox{DE-FG02-87ER40365}. This work contributes to the aims of the U.S. Department of Energy ExoHad Topical Collaboration, contract DE-SC0023598.
 \end{acknowledgments}

\bibliographystyle{apsrev4-2}
\bibliography{references} 
\end{document}